# Users' Prompting Strategies and ChatGPT's Contextual Adaptation Shape Conversational Information-Seeking Experiences

**Haoning Xue[1], Yoo Jung Oh[2], Xinyi Zhou[3], Xinyu Zhang[2], Berit Oxley[4]**

[1]University of Utah
[2]Michigan State University
[3]Boise State University
[4]Emory University

haoning.xue@utah.edu, ohyoojun@msu.edu, xinyizhou@boisestate.edu, zxinyu@msu.edu, berit.oxley@emory.edu

## Abstract

Conversational AI, such as ChatGPT, is increasingly used for information seeking. However, little is known about how ordinary users actually prompt and how ChatGPT adapts its responses in real-world conversational information seeking (CIS). In this study, a nationally representative sample of 937 U.S. adults engaged in multi-turn CIS with ChatGPT on both controversial and non-controversial topics across science, health, and policy contexts. We analyzed both users' prompting strategies and the communication styles of ChatGPT's responses. The findings revealed behavioral signals of digital divide: only 19.1% of users employed prompting strategies, and these users were disproportionately more educated and Democrat-leaning. Further, ChatGPT demonstrated contextual adaptation: responses to controversial topics contain more cognitive complexity and more external references than to non-controversial topics. Notably, cognitively complex responses were perceived as less favorable but produced more positive issue-relevant attitudes. This study highlights disparities in user prompting behaviors and shows how user prompts and AI responses together shape information-seeking with conversational AI.

## Introduction

Conversational AI systems powered by Large Language Models (LLMs), such as ChatGPT, Claude, and Gemini, have empowered existing conversational information-seeking (CIS) systems and become critical information sources for critical societal issues[1]. Among these conversational AI systems, ChatGPT remains the most widely used AI tool, potentially affecting millions of users[2]. A recent Pew Research report highlights that Americans are increasingly turning to ChatGPT for new information[3]. Users increasingly use and trust such systems for information on important societal issues across health, science, and public policies, even more than traditional search engines like Google[4], because conversational AI is often perceived as a nonpartisan, objective information source (e.g., algorithm appreciation[5] and machine heuristics[6]).

This growing reliance on conversational AI raises questions about the communication dynamics between users and AI systems. On the user side, how users actually prompt represents behavioral signals of AI literacy in action and the second-level digital divide in interacting with conversational AI, beyond self-reported attitudes and behavioral intentions[7]. The first-level digital divide revolves around technology access and adoption, while the second-level digital divide[7] zooms in on the disparities in users' skills in using technologies. Existing human-centered studies have mostly examined broader patterns of first-level digital divide in AI perception, adoption, and trust, showing how disparities in adopting conversational AI exist across income[8] and education levels[9]. Further, factors such as information sufficiency[10], risk perceptions[11], and social support[12] drive information-seeking intentions with conversational AI as well. Research shows that prompting strategies such as asking AI to adopt a persona[13] and requesting step-by-step reasoning[14] can impact the accuracy, quality, and style of AI responses. Yet little is known about how ordinary users actually use these strategies in real-world information seeking. Prompting as a behavioral signal of AI literacy in action and the second-level digital divide has been largely overlooked.

On the AI system side, how AI responds is equally important. Existing research on CIS has primarily focused on content-related issues[15], such as accuracy, bias, and hallucination. But we argue that communication styles are central to human-AI interactions[16], particularly in CIS contexts, where the interaction is dynamic, personalized, and iterative. In these dyadic information-seeking exchanges, AI responses' tone, structure, and style are critical in shaping user experiences, influencing attitude formation and change, and facilitating the exchange of ideas and factual information[17]. This is more pressing for controversial topics such as vaccination, climate change, and immigration, which are often polarized and emotionally charged. Existing work has been focusing on content-level adaptation, finding that ChatGPT's responses on controversial issues are often more comprehensive[18] and may reinforce users' confirmation biases[19]. However, the extent of such contextual adaptation in communication styles and its downstream effects on user perceptions remains unclear.

Taken together, both how users prompt and how AI responds in CIS remain understudied. This study zooms in on ChatGPT, the most widely used conversational AI[2] and addresses these research gaps by examining three interconnected levels of human-AI interactions in CIS: (1) how users adopt prompting strategies, (2) how ChatGPT's communication styles adapt to user prompts and context controversy, and (3) how ChatGPT's communication styles shape users' AI perceptions and context-relevant attitudes.

## Prompting Strategies in Conversational Information-seeking

Prompting strategies refer to user inputs with specific instructions or structures that can optimize AI responses[20]. While traditional information-seeking research focuses on search tactics and queries in online databases and search engines[21], user prompts are natural language inquiries in dialogues. Examples include asking AI to adopt a persona[13], requesting step-by-step reasoning[14], or specifying response structures and styles[22].

Much of the research on prompting engineering has focused on the AI system side, testing how prompt engineering may improve LLM performance. However, how ordinary users actually adopt these prompting strategies in everyday information-seeking remains largely unclear. A recent study on Google Bard found that fewer than 6% of real-world prompts showed evidence of prompt engineering[23]; these advanced strategies were rarely used in practice. Therefore, we seek to understand the prevalence of 8 prompting strategies (e.g., adopt a persona, step-by-step reasoning) in CIS, according to ChatGPT prompting guidelines[24].

**RQ1.** How do users use prompting strategies for information seeking with ChatGPT?

Beyond this gap between prompting engineering research and how users actually prompt, another question concerns *who* actually adopts prompting strategies. The digital divide exists not only in AI access and usage but also in AI skills[8,9]. Prior research shows that individuals with higher family socioeconomic status[25] tend to have higher AI literacy, are more likely to use AI for work and education, and use more abstract language in AI conversations[26]. In contrast, less educated users reported having worse experiences with ChatGPT in seeking information on controversial issues such as climate change[27]. Building on prior findings of structural inequalities in accessing and using AI, we zoom in on the socio-demographic disparities in using prompting strategies, as prompts provide behavioral signals of AI literacy in action beyond self-evaluations and perceptions[28].

**RQ2.** How are users' prompting strategies associated with socio-demographics?

## Communication Styles Adaptation in Conversational Information-seeking

Communication styles are central to how people shape perceptions and form attitudes. Affinity-seeking language increases social likability in conversations[29]; intensive and extreme language enhances persuasion[30]. Beyond static linguistic features, the dynamic exchange of communication styles is equally important[31]. Reciprocity of communication styles improves interpersonal relationship satisfaction[32]; in online support groups, the alignment of positive and negative sentiments increases future support-seeking behaviors[33].

Conversational AI exhibits similar adaptive behaviors by mirroring linguistic cues and tones of user inputs[34]. But compared with human conversations, conversational AI's adaptation is often asymmetrical, with AI accommodating users more[35]. Such adaptation can improve conversation flow and satisfaction[36], but it can introduce risks when conversational AI overly adapts, such as excessively agreeing with users (a phenomenon known as "sycophancy"[37]) or treating health-related claims as true by default[38], which is parallel to humans' truth-default bias[39]. This line of work confirms conversational AI's adaptability, but how such adaptation applies to users' prompting strategies is unclear.

This adaptation extends to issue contexts and stances, with both similar (i.e., reciprocity) and dissimilar (i.e., compensation) styles [40]. Overall, ChatGPT's responses on controversial issues are often more comprehensive[18] and liberal-leaning[41]. When discussing controversial issues, ChatGPT compensates negatively-toned prompts with responses in a neutral or positive tone[42]. However, contradictory evidence exists that ChatGPT exhibits more hostility and higher emotional intensity in discussing controversial issues[43]. The mixed evidence raises questions about how conversational AI adapts to issue controversy consistently in natural conversations with users.

**RQ3.** How do (a) issue controversy and (b) users' prompting strategies influence ChatGPT's responses in information-seeking interactions?

Research shows that conversational AI's ability to adapt and customize communication styles influences how users perceive human-AI interaction and shapes attitude and behavior change[44]. For example, conversational AI matching users' linguistic styles has been shown to increase donations to charities[45]. Elaborateness and politeness in Amazon Alexa's responses improve user satisfaction[46]. Conversational AI's adaptation to users' communication styles improves the overall conversation flow, understanding, and satisfaction[36,47]. Although empirical evidence reveals the persuasiveness of conversational AI in persuasion contexts, it is unclear how AI responses, especially communication styles, may influence users' AI perceptions and issue-specific attitudes. Therefore,

**RQ4.** How do ChatGPT's responses influence users' (a) AI perceptions and (b) issue-specific attitudes?

Furthermore, while prior work primarily examines generic linguistic features such as functional words and sentiment, we argue that focusing on specific communication styles more relevant to CIS interactions provides deeper and more meaningful insights. Therefore, in this study, we emphasize five information-related communication styles[48,49]. These communication styles go beyond the content exchanged; instead, these styles reflect how information is presented in both user prompts and AI responses. It is important to note that while we use information-seeking to refer to the broader context of users seeking information from conversational AI, information-seeking below captures question-asking language styles.

- **Self-revealing**: sharing one's own experience and opinion
- **Information-seeking**: asking questions or seeking advice
- **Fact-oriented**: discussing or providing factual information
- **Action-seeking**: calling for action or giving actionable advice
- **Cognitive complexity**: expressing or explaining complexly organized concepts

To answer these questions, we conducted a 3 (issue: health, science, policy) × 2 (issue controversy: controversial, non-controversial) between-subject experiment. A nationally representative sample of 937 participants = engaged in a conversation with ChatGPT to seek issue-specific information. We highlight three main findings in this study. First, 19.1% of users have used at least one prompting strategy, predicted by education level and party affiliation. Second, issue controversy drives ChatGPT responses' communication styles (e.g., action-seeking, cognitive complexity) and the number of citations, while the communication styles of user prompts remain consistent across issue controversy. Lastly, ChatGPT responses' cognitive complexity influences users' AI perceptions; users perceived more cognitively complex AI responses as less favorable but experienced more positive issue-relevant attitude changes with cognitively complex AI responses, suggesting the implicit influence of cognitive complexity on attitude changes.

# Methods

## Experiment Design and Procedure

This between-subject experiment has a 3 (issue: health, science, policy) × 2 (issue controversy: controversial, non-controversial) (see Figure 1 for experimental procedure). All study procedures have been approved by the university's Institutional Review Board. A nationally representative sample of U.S. adults (n = 937) was recruited on Prolific in November 2024.

**Figure 1.** Flow diagram of experiment procedure.

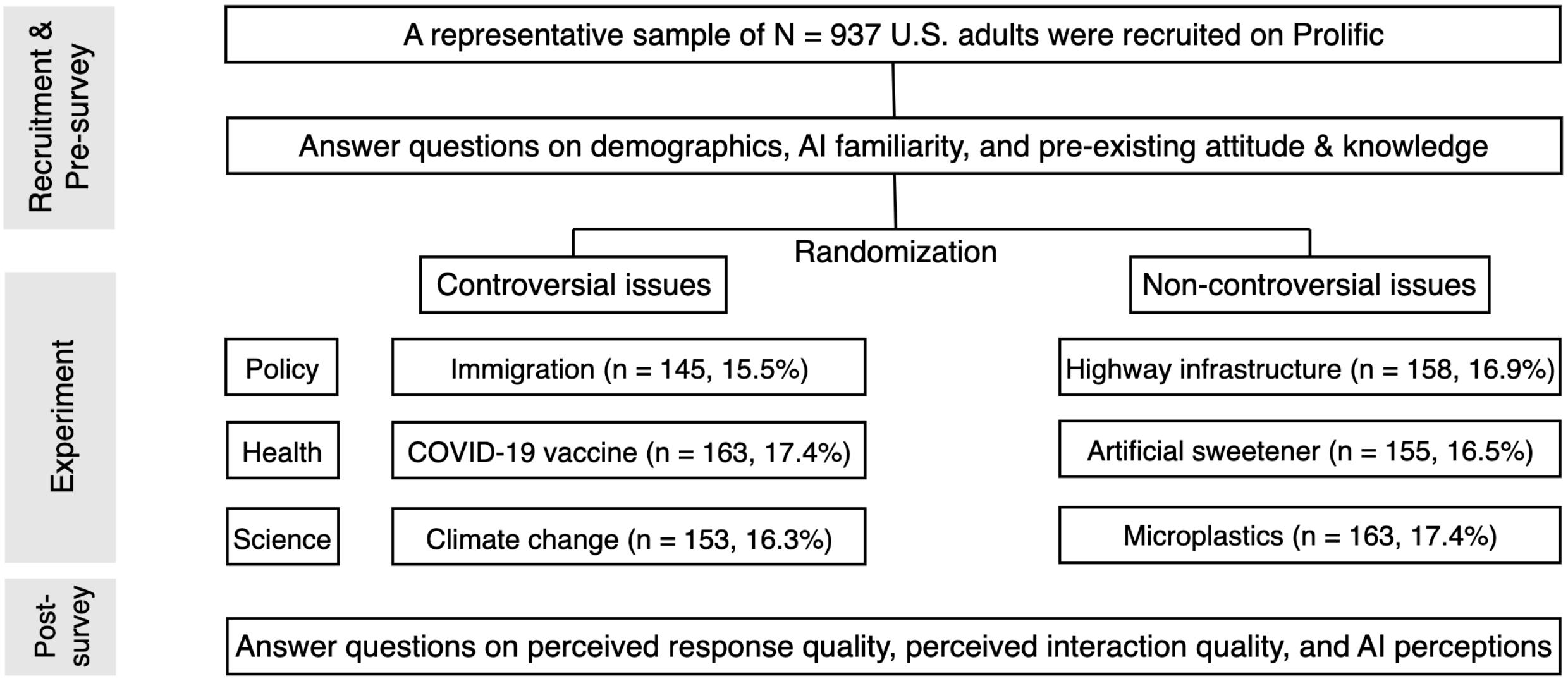

Upon providing consent, participants reported demographics and pre-existing attitudes on AI and 6 issues. Afterward, participants were randomized to one of the six topics across issue controversy and topics in Figure 1. Participants were instructed to seek issue-relevant information from ChatGPT for a hypothetical social scenario. Participants were required to interact with GPT-4o for at least 5 turns before submitting the shareable conversation link for compensation, which is meant to ensure sufficient engagement and meaningful interactions[27]. See Figure 2 below for an example instruction; see Table S1 in the Supplemental Information for all instructions. Lastly, participants reported their perceptions of AI responses, interaction experience, and post-experiment issue-specific attitudes. Among the 937 participants, 473 were female (50.8%), with a mean age of 45.1 years, 70.0% white, Democrat-leaning (55.7%), and median annual household income between \$50,000 and \$74,999. See Table S2 for demographics summary. The median participation time was 18 minutes.

**Figure 2.** Illustration of example information-seeking instruction and interaction.

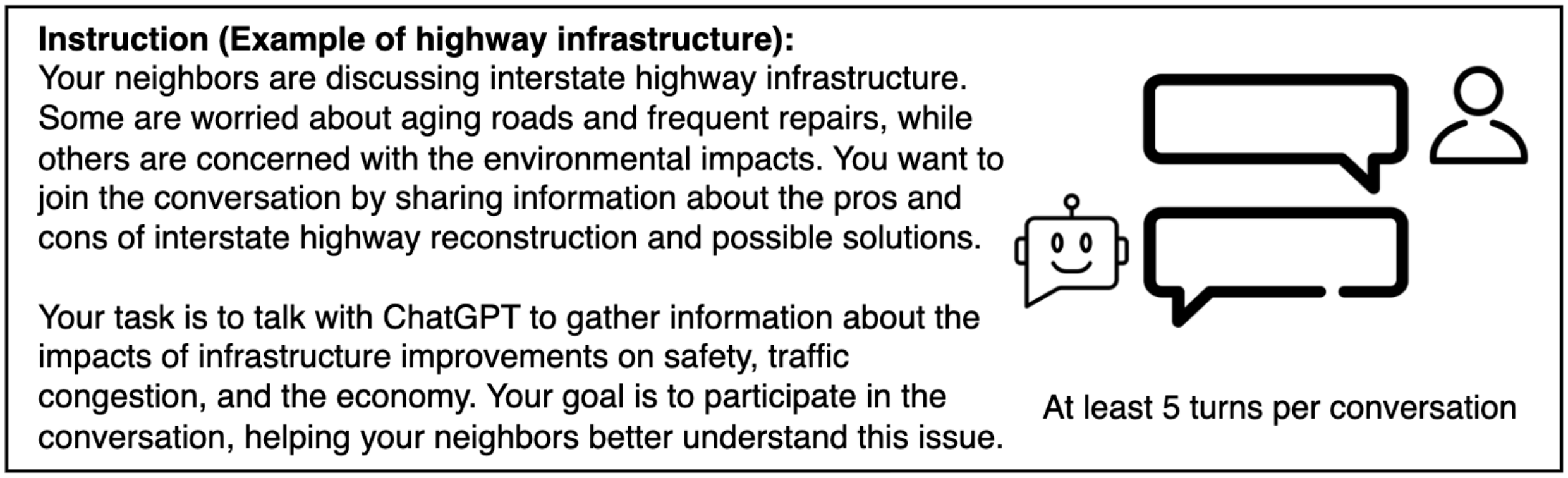

*Manipulation check.* To ensure the manipulation of issue controversy is successful, participants rated how controversial the six topics are on a 5-point scale. Participants perceived immigration (M = 4.36, SD = 0.83) as the most controversial and highway infrastructure as the least (M = 1.96, SD = 0.96). On average, three controversial topics (M = 3.78, SD = 0.73) were perceived as more controversial than non-controversial topics (M = 2.29, SD = 0.76; t = 29.81; $p < .001$).

## Conversation Data Collection

We retrieved 747 valid URLs from 937 responses. We excluded n = 137 inaccessible links (e.g., broken chat links, chat links with sharing turned off) and n = 53 irrelevant conversations (i.e., chats irrelevant to the study context) from further analysis. We scraped all conversations and related metadata, including conversation titles, the ChatGPT models used, and external links included in ChatGPT's responses. On average, there were 6 turns per conversation (M = 5.56, SD = 1.98, Min = 2, Max = 44,

Median = 5). Most chats (n = 722, 96.7%) were conducted with GPT-4o as instructed, while 24 chats (3.2%) used GPT-4o-mini and 1 chat (0.1%) used GPT-4. We included relevant conversations that did not strictly follow instructions, as they still reflect natural CIS with ChatGPT.

## Communication Style Extraction

We used the Symanto Psychology API[48] and Linguistic Inquiry and Word Count (LIWC) 2022[50] to extract 5 communication styles related to information seeking. See Table 1 below for definitions, summary statistics, and examples. For every conversation, we aggregated user prompts and ChatGPT responses across conversation turns and calculated communication styles, respectively. This is meant to capture an overall interaction pattern while reducing noise arising from turn-to-turn variability. First, we used the Symanto Psychology API[48], which is based on a fine-tuned BERT model, to extract communication styles related to self-revealing, information-seeking, fact-orientation, and action-seeking. Each of the four communication styles is assigned a probability score between 0 and 1, with higher values indicating a stronger presence in the text. Second, we used LIWC to calculate the categorical-dynamic index (CDI) to capture cognitive complexity[49]. CDI reflects the extent to which a text expresses complexly organized concepts, which can be inferred with functional words. We normalized CDI for comparison across communication styles, with higher values indicating more cognitive complexity.

**Table 1.** Definitions, summary statistics, and examples of communication styles.

| Communication style | Author | M | SD | Example |
|---|---|---|---|---|
| Self-revealing: Speakers share personal information or experiences. | User | 0.39 | 0.46 | I love stevia and use it primarily but I need some convincing information to tell others how awful artificial sweeteners are. |
| | ChatGPT | 0.06 | 0.23 | Of course, Neighbor! I'd really appreciate any insights you have on the reconstruction. Better safety, less traffic during rush hours, and a positive impact on the economy all sound like great outcomes. Please, go ahead and tell me more! |
| Information-seeking: Direct or indirect questions searching for information. | User | 0.97 | 0.17 | Are all the new boosters out now? When should they receive them for best holiday immunity? |
| | ChatGPT | 0.07 | 0.24 | I'd be happy to help with any questions you have about climate change. What would you like to know? |
| Fact-oriented: Factual and objective statements. | User | 0.11 | 0.30 | Why do you defend illegal immigration? The act of crossing the border illegally is a crime in and of itself. |
| | ChatGPT | 1.00 | 0.04 | The side effects of the COVID-19 vaccines are generally mild and temporary. Here's a breakdown of the most common ones. |
| Action-seeking: Direct or indirect requests, suggestions, and recommendations expecting action from other people. | User | 0.10 | 0.29 | Can you put yourself into the shoes of someone against immigration and tell me your points and try not to bias what I've said at all |
| | ChatGPT | 0.14 | 0.34 | 1. That's a great initiative! Here are some ways to make it engaging and practical for your neighbors: 1. **Host an Informative Event or Workshop**<br>2. Investing in a proper filtration system is the best way to ensure you're effectively reducing microplastic exposure. Would you like more details on any of these methods? |
| Cognitive complexity: Expressing or explaining complexly organized concepts | User | 0.56 | 0.15 | Summarize the impact of immigration on the US, Include topics like national security and economy. |
| | ChatGPT | 0.59 | 0.10 | Public opinion on immigration in the United States is notably divided, often along partisan lines. Here are some key statistics illustrating these divisions. |

## Factual and Structural Feature Extraction

In addition to communication styles, we extracted two distinctive features capturing the factual and structural dimensions of ChatGPT responses.

*Citation count.* Whether solicited or not, the number of citations provided in AI responses signals credibility and trustworthiness[51]. We extracted the number of external links in every ChatGPT response and then averaged these counts for each conversation, accounting for differences in conversation duration. This yields the average citation count to represent the factual basis of AI responses (Min = 0, Max = 12, M = 0.62, SD = 1.66).

*Structure count.* Structure count is used to represent the readability and clarity of ChatGPT responses, as structured elements such as headings and bullet points can enhance comprehension and facilitate learning[52]. Structure count (Min = 0, Max = 104.4, M = 21.89, SD = 14.23) is the sum of the average number of structural elements in ChatGPT responses for every conversation, including the number of headings (Min = 0, Max = 12, M = 2.55, SD = 2.36), bullet points (Min = 0, Max = 32.4, M = 6.78, SD = 5.65), numbered points (Min = 0, Max = 12, M = 1.92, SD = 2.00), and bolded texts (Min = 0, Max = 64.8, M = 10.63, SD = 7.24).

## Prompting Strategies Annotation

Referring to ChatGPT prompting guidelines[24], we identified 8 prompting strategies and grouped them into three broad categories based on whether users provided or requested information: user-supplied information, style-related requests, and content-related requests (see Table 2 below for details). To systematically identify prompting strategies, we used ChatGPT to annotate all user prompts (n = 4,154; see Table S3 for the annotation instructions). We chose this approach to capture how ChatGPT perceives and interprets the prompting strategies that users employ, which may directly influence how ChatGPT formulates its responses. As an additional validity check, three trained coders independently validated a random subset of n = 884 (21.3%) user prompts, achieving high inter-coder reliability (IRR = .814). Further, we calculated the weighted F1 score to account for data imbalance. The weighted F1 score of 0.903 shows that ChatGPT's annotation has a high accuracy and a high consistency with human interpretation.

**Table 2.** Summary statistics of prompting strategies with example prompts.

| Prompt Category | Prompt Strategy | #users (%) | #prompts (%) | Example |
|---|---|---|---|---|
| User-supplied information | Provide delimiter | 3 (0.4%) | 3 (0.1%) | what are some specific solutions that involve policy change rather than """reusable straws""" ideas. |
| | Provide example | 3 (0.4%) | 4 (0.1%) | I mean... like... something we all do but accept the risk where the risk is much higher... like, oh, driving to Kroger, for example. What else is very common but with a much higher statistical risk than getting the covid shot? |
| | Provide context | 75 (10.0%) | 87 (2.1%) | My friends have been discussing the health effects of artificial sweeteners. I want to appear informed. I want to help my neighbors understand the safety and potential health risks associated with artificial sweeteners. |
| | **Sum** | **78 (10.4%)** | **94 (2.3%)** | |
| Style-related requests | Specify length | 22 (2.9%) | 23 (0.6%) | Summarize in 100 words or fewer how to convince someone that the new Covid-19 booster is worth getting. |
| | Specify style | 46 (6.2%) | 54 (1.3%) | Could you answer that question in a more briefly and in a more conversational tone? |
| | **Sum** | **67 (9.0%)** | **77 (1.9%)** | |
| Content-related requests | Specify persona | 5 (0.7%) | 8 (0.2%) | I want you to act like an economics expert and tell me the pros and cons of the current US immigration policy. |
| | Request references | 54 (7.2%) | 68 (1.6%) | Please cite sources for medical information on artificial sweeteners. |
| | Request stepwise reasoning | 9 (1.2%) | 11 (0.3%) | Can you break this down into steps we should take to reverse global warming? |
| | **Sum** | **67 (9.0%)** | **87 (2.1%)** | |

## Self-Reported Measures

Tables S4-S8 list question items for measures below. See Table S9 for summary statistics of key variables by issue topic.

*Demographics*. We asked participants to report their age, gender, race, education, income, and political affiliation. Gender and race were converted to binary variables (gender: 1 for female, 0 for male; race: 1 for white, 0 for non-white). Political affiliation was measured by one question on participants' political party affiliation on a 7-point scale (1: a strong Democrat, 7: a strong Republican) ($M = 3.29$, $SD = 2.02$).

*Issue-specific pre-existing attitude and knowledge*. Participants' pre-existing attitudes were measured with one question per issue on a 4-point scale, such as *How effective do you think the COVID-19 vaccines are?* Further, participants' pre-existing knowledge was measured with one question: *How would you rate your level of knowledge about the following topics?*

*AI familiarity*. Users' familiarity with AI technologies[53] was measured by one question: *Which of the following technologies, if any, uses artificial intelligence (AI)?* Responses included 12 AI products (e.g., Google search, chatbots). The number of AI products identified indicates familiarity with AI (Min = 0, Max = 12, M = 8.55, SD = 3.08).

*Issue-specific attitude (post-experiment)*. After interacting with ChatGPT, participants reported their issue-specific attitudes again for the assigned issue, using four questions per issue on a 5-point scale. Specifically, these questions are about COVID-19 vaccine safety (M = 3.91, SD = 1.15, Cronbach's α = .93), climate change severity (M = 4.07, SD = 0.96, Cronbach's α = .90), immigration benefits (M = 3.59, SD = 1.08, Cronbach's α = .93), artificial sweetener safety (M = 2.70, SD = 1.03, Cronbach's α = .88), microplastics severity (M = 4.04, SD = 0.84, Cronbach's α = .92), and highway reconstruction benefits (M = 3.89, SD = 0.64, Cronbach's α = .60).

*Perceived AI response quality*. Participants rated AI response quality with 12 items on a 7-point scale[54], such as *generic–in-depth,* and *clear–ambiguous* (M = 5.72, SD = 0.89, Cronbach's α = .89).

*Perceived AI interaction quality*. Participants rated the quality of the information-seeking interaction with AI with 7 items on a 7-point scale[54], such as *In this conversation, ChatGPT was able to understand my questions and instructions clearly* (M = 4.29, SD = 0.55, Cronbach's α = .73).

*AI perceptions: perceived likability, trustworthiness, and intelligence*. Participants evaluated AI with a source evaluation measure on a 7-point scale[55], including perceived likability (M = 5.70, SD = 1.18, Cronbach's α = .90), trustworthiness (M = 5.56, SD = 1.11, Cronbach's α = .92), and intelligence (M = 6.03, SD = 1.06, Cronbach's α = .90).

## Analysis Plan

To answer RQ1 about the prompting strategies used, we provide descriptive and summary statistics of the prompting strategies used. To answer RQ2 about the relationship between socio-demographics and users' prompting strategy usage, we conduct a negative binomial regression with the number of prompting strategies used as the dependent variable. Negative binomial models are appropriate for the overdispersed prompting strategy count data (dispersion = 1.40, $p < .001$). We include gender, age, race, education, income, and political affiliation as independent variables with five covariates (i.e., issue controversy, issue topic, pre-existing attitude, pre-existing knowledge, and AI familiarity) controlled.

To answer RQ3 about the influence of issue controversy and users' prompting strategies on ChatGPT responses, we conduct linear regression models with the citation count, structure count, and five communication styles, respectively. Issue controversy and prompting strategy count (pure count and by three categories in separate models) are included as independent variables. Further, issue topic, users' pre-existing attitude, user prompts' five communication styles, and word counts of user prompts and ChatGPT responses are controlled as covariates.

To answer RQ4 about the effects of ChatGPT responses on AI perceptions and issue-specific attitudes, we run similar linear regression models with AI perceptions and issue-specific attitudes as dependent variables, respectively. We include ChatGPT responses' citation count, structure count, and five communication styles as independent variables. Users' demographics, pre-existing attitudes, AI familiarity, and issue controversy are controlled as covariates.

# Results

## Prompting Strategies are Rarely Used, Mostly by Educated and Democrat-leaning Users

RQ1 asked about how users adopt prompting strategies in CIS. We found that prompting strategies were rarely used in information seeking (see Table 2 above for descriptive statistics). Across all conversations, 19.1% users (n = 179) used at least one prompting strategy, resulting in n = 258 prompts (6.2% of all user prompts). Most users employed straightforward, plain questions rather than strategically engineered prompts in information-seeking with ChatGPT.

Among prompts with any prompting strategies, the most commonly used strategies were providing contextual information (n = 87, 2.1%), requesting external references (n = 68, 1.6%), and specifying ChatGPT's response styles (n = 54, 1.3%). In contrast, the least frequently used strategies were providing delimiters (n = 3, 0.1%), providing examples (n = 4, 0.1%), and requesting ChatGPT to adopt a persona (n = 8, 0.2%). It is worth noting that the high frequency of providing contextual information might be inflated by the experimental setting of information seeking for a hypothetical neighborhood discussion, as users tended to repeat back to ChatGPT as additional contextual information. Similarly, while providing examples is common for task-oriented prompting (e.g., few-shot prompting), it is less relevant in CIS.

For RQ2 about the association between socio-demographics and the usage of prompting strategies, we found that education and political affiliation were significant predictors (see Figure 3 below for visualization; see Table S10 for full regression models). More educated and Democrat-leaning users employed more prompting strategies, especially content-related requests. Further, users with higher levels of AI familiarity used more style-related requests, suggesting that users with more AI knowledge are more likely to fine-tune the styles of ChatGPT responses. Females were less likely to use content-related requests, while older users were less likely to prompt with user-supplied information. In addition to socio-demographics, users' pre-existing knowledge negatively predicted prompting strategy usage. It is consistent with the notion that more knowledgeable users are less motivated to seek information. Overall, prompting strategies were not evenly adopted across the population, but skewed toward the educated and Democrat-leaning users.

**Figure 3.** Regression coefficients and confidence intervals predicting the count of (A) prompting strategy and three prompting strategy categories, respectively: (B) user-supplied information, (C) style-related requests, and (D) content-related requests. Non-significant regression coefficients are marked in gray. See Table S5 in Supplemental Information for full regression tables.

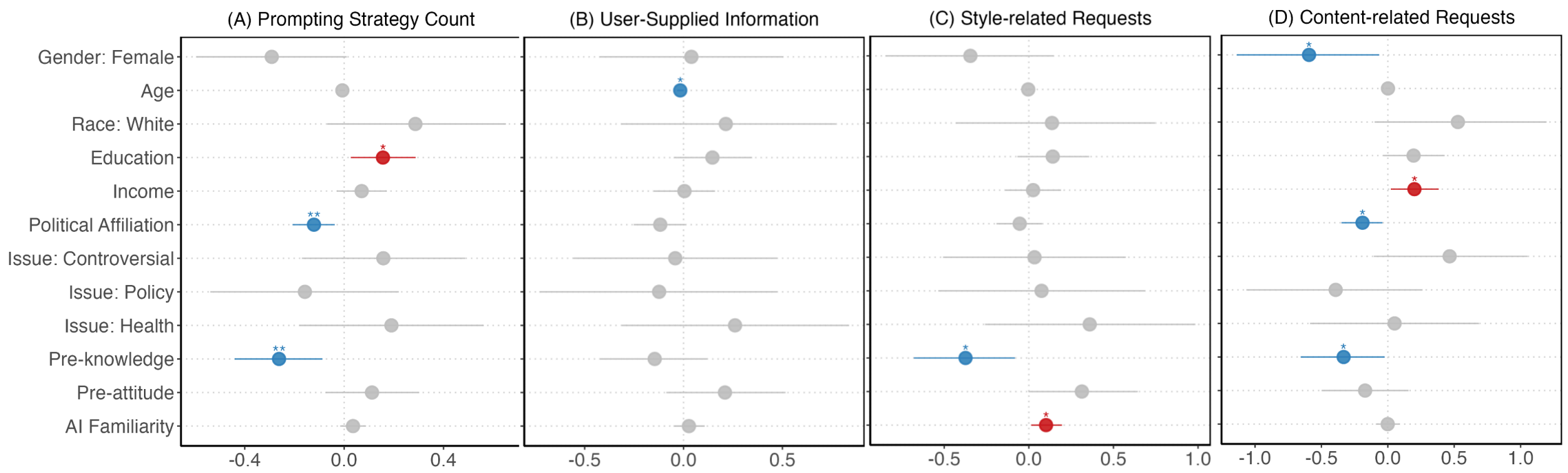

## ChatGPT's Responses Adapt to Controversial Issues and Users' Prompting Strategies, Especially in Terms of Cognitive Complexity

RQ3 asked about how issue controversy and users' prompting strategies influence ChatGPT's responses. Figure 4 below reports regression coefficients for both sets of models: one using the overall number of prompting strategies (Panels A1-A2), and another using three prompting categories as predictors (Panels B1-B4). Consistently, for controversial issues, ChatGPT provided more references to external links, generated less structured responses, and used more cognitively complex and action-oriented language (Figure 4 Panels A1, B1). These effects held when users' communication styles were controlled, suggesting that ChatGPT adjusts its communication styles to contextual cues rather than simply mirroring user prompts.

Further, the prompting strategy count overall did not significantly predict ChatGPT responses (Figure 4 Panel A2), but more nuanced patterns emerged with three prompting strategy categories (Figure 4 Panels B2-B4). When users provided additional information, ChatGPT responded with more action-oriented language. When users made style-related instructions, ChatGPT toned down cognitive complexity and used simpler language. However, when users made content-related requests, ChatGPT used more cognitively complex language for elaborate reasoning. These findings suggest that different types of prompting strategies activate distinct communication styles in ChatGPT responses. Overall, cognitive complexity is the most sensitive communication style, influenced by both user prompts and issue controversy.

**Figure 4.** Regression coefficients and confidence intervals for the effects of issue controversy and prompting strategies. Panels A1-A2 show results from models with issue controversy and prompting strategy count as independent variables. Panels B1-B4 show results with issue controversy and three prompting strategy categories (i.e., user-supplied information, style-related requests, and content-related requests). Non-significant regression coefficients are marked in gray. Covariates were controlled in all models. See Tables S11-S12 for full regression tables.

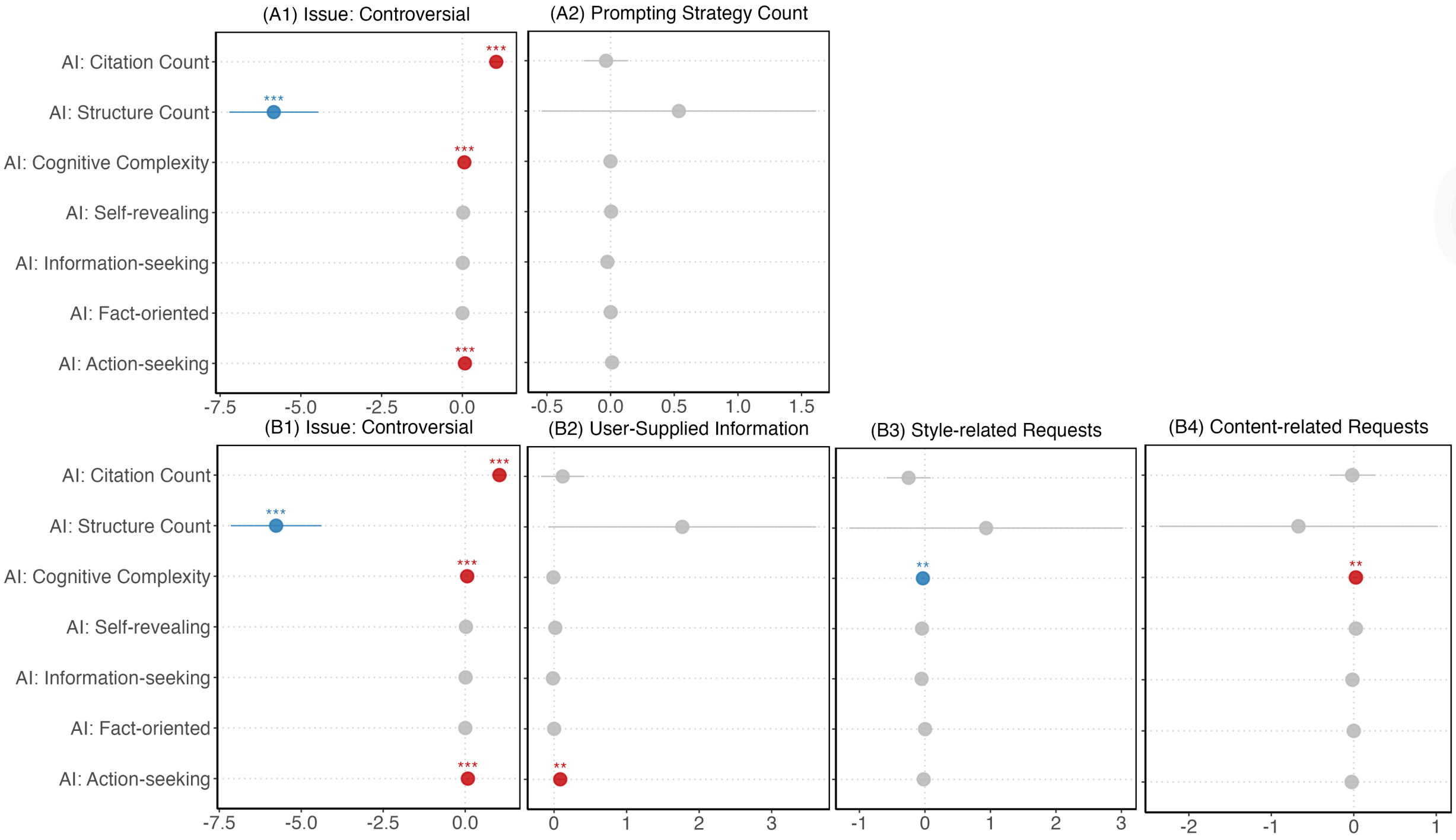

## Cognitively Complex ChatGPT Responses Decreased Users' AI Perceptions but Improved Issue-Specific Attitudes

RQ4 asked about how ChatGPT responses may impact users' AI perceptions and issue-specific attitudes. Across all communication styles, only cognitive complexity showed significant impacts (see Figure 5 below for illustration). ChatGPT with higher cognitive complexity was perceived as less favorable, with lower levels of perceived response quality, likability, and intelligence. This pattern suggests that users prefer straightforward responses with plain language in CIS. Meanwhile, cognitively complex responses positively affected issue-specific attitudes. Specifically, controlling for users' pre-existing attitudes, cognitively complex responses resulted in stronger risk perceptions of microplastics and climate change, as well as increased support for vaccination, artificial sweetener, immigration, and highway reconstruction. This finding suggests that while users may perceive complex ChatGPT responses as less favorable, the cognitively demanding communication style nonetheless contributed to the persuasiveness of ChatGPT responses, implicitly influencing issue-relevant post-interaction attitudes with one-shot interaction.

**Figure 5.** Regression coefficients and confidence intervals for the effects of cognitive complexity in ChatGPT responses on users' AI perceptions and issue-specific attitude, with covariates controlled. Non-significant regression coefficients are marked in gray. See Table S13 for full regression tables.

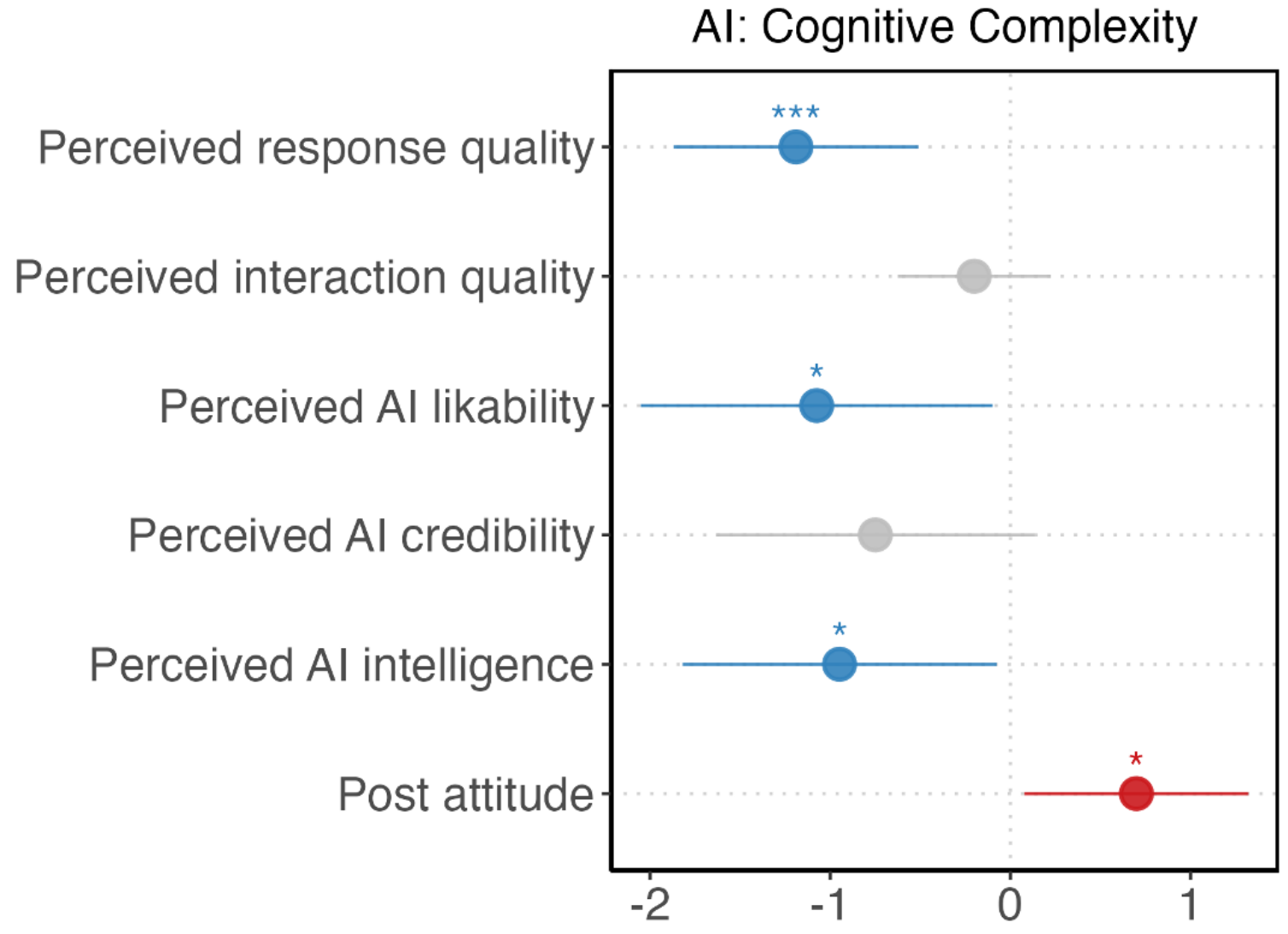

## Discussion

With a between-subjects experiment of conversational information-seeking and analysis of real conversations, our study highlights three key findings. First, people rarely use prompting strategies for CIS, and these strategies are mostly used by educated and Democrat-leaning individuals. Second, the communication styles of ChatGPT's responses adapt to users' prompting strategy and issue controversy. Lastly, the cognitive complexity of ChatGPT decreased users' AI perceptions while positively shaping issue-specific attitudes. Below, we interpret potential explanations for our findings, propose actionable implications for conversational AI system design, and suggest directions for future research.

First, we found that only 19.1% of users employed at least one prompting strategy, and these users were disproportionately educated and Democrat-leaning. This finding suggests a second-level digital divide, such that disparities exist beyond technology adoption but in how people actually use technologies[7]. It is consistent with previous findings that education predicts the digital divide[56]. In addition, it aligns with prior work on partisan gaps in technology adoption: Democrats are more likely to use social media[57] and are more open to using AI in policy-making[58] than Republicans. These findings reveal how sociodemographics and ideological leaning influence how people approach and perceive emerging technologies.

This finding fills a critical gap in the research on conversational AI that focuses on system-side disparities in AI responses by highlighting human-side disparities and demonstrating how users actually prompt for information seeking. At the same time, the relatively low overall use of prompting strategies suggests that advanced prompting may not be essential for CIS: systems like ChatGPT can already interpret user intent and sustain natural conversations without engineered inputs[59]. Future research, especially surveys and qualitative studies, is needed to understand when and why users adopt prompting strategies.

Second, our results suggest that ChatGPT adapts communication styles to not only users' prompting strategies but also issue controversy, demonstrating contextual adaptability. We found that different types of prompting strategies activated distinct communication styles, which suggests that ChatGPT can adjust both content and styles to user intent. For example, when users employed content-related prompting strategies such as stepwise reasoning, ChatGPT prioritized response depth and elaborated with higher cognitive complexity.

Moreover, ChatGPT's responses to controversial issues contain more action-oriented language, higher levels of cognitive thinking, less structured responses (e.g., bullet points, subheadings), and more external references in comparison with non-controversial issues. It suggests that ChatGPT also adapts to broader social contexts in terms of communication styles. This finding adds to the growing body of research on AI's social intelligence, which primarily focuses on adaptation to human users[60]. While this contextual adaptability does not necessarily imply that AI is socially aware or socially intelligent, it indicates that AI systems can resemble human-like adaptability by recognizing and responding to unique communication patterns across issue controversy in the training data. For instance, human conversations on controversial issues tend to be more analytical and emotional[61]; ChatGPT may reproduce similar linguistic patterns. At the same time, given the existing finding that AI may amplify confirmation bias[19], such contextual adaptation needs further academic attention to prevent potential harm caused by AI systems.

Lastly, our results reveal a seemingly contradictory finding: users reported more negative AI perceptions but more positive issue-relevant attitudes with cognitively complex ChatGPT responses, with pre-existing attitudes and issue controversy controlled. Cognitive complexity may have implicit persuasive effects, even if users consciously dislike it. It aligns with prior work on ChatGPT-3, where opinion-minority users reported worse experiences but still shifted their attitudes, potentially due to cognitive dissonance[27]. One possible explanation lies in the cognitive disfluency, which refers to the mental difficulty of processing cognitively complex responses[62]. Complex information may disrupt cognitive fluency and decrease positive affect[63], but it can also activate systematic processing and facilitate implicit learning[64]. Therefore, while users may prefer simple responses, they may unconsciously allocate cognitive resources to systematically process cognitively complex AI responses, thereby having more positive context-relevant attitudes.

Our findings suggest that conversational AI should be designed to increase accessibility and inclusivity. First, because most people do not use advanced prompting, systems should support effective interactions without assuming technical expertise. In addition, ensuring that conversational AI adapts in ways that promote constructive dialogue, rather than reinforcing existing biases, will be critical as these tools are increasingly used in everyday contexts. Second, our findings point to a potential design challenge: cognitively complex responses strengthen issue-specific attitudes but reduce how favorably users view the AI. To reconcile this tension, systems could present complexity in more user-friendly ways. For instance, it would be beneficial to combine complex reasoning with stylistic elements, such as conversational tone or concrete examples, to maintain likability even when information is dense. This is essential to build credible, likable, and accessible conversational AI systems.

Several limitations of this study need to be noted. First, the information-seeking interactions with ChatGPT in this experiment were not fully controlled. Although users were instructed to interact with GPT-4o, a few participants did not follow instructions. Further, we did not control the alternative settings, such as memory functions, which could influence the conversation quality and experience. Although this design was meant to capture naturalistic CIS with ChatGPT, future studies should aim for a cleaner experimental design, ensuring standardized conversation settings to minimize variability in user interactions. Second, this study focuses exclusively on ChatGPT. While this decision allows for consistency in experimental conditions, it may limit the generalizability of our findings to other conversational AI systems that exhibit different communication styles or interaction patterns. Future research should replicate this study across multiple conversational AI systems, such as Gemini and Copilot, to assess the variation of contextual and individual-level adaptations across AI systems. Lastly, our findings on communication styles and their persuasive effects are based on a one-shot interaction, and we did not measure attitude change directly. Given that users often engage in conversations with AI over time, future research should expand the scope of the current study and investigate the longitudinal effects of communication styles on users' perceptions and attitudes with repeated conversations and information exposures.

**Supplemental Information for**

**Users' Prompting Strategies and ChatGPT's Contextual Adaptation Shape Conversational Information-Seeking Experiences**

**Table S1.** Study instructions across issue contexts.

| Controversy | Issue | Instructions |
|---|---|---|
| Controversial | Health | **[COVID-19 booster]**<br>COVID-19 cases are surging this summer, and a new COVID-19 booster has just been released. Your neighbors are concerned and thinking about getting the booster, but they aren't sure how effective it is. They've heard mixed information and are skeptical about whether getting the booster is worth it. You want to join the conversation by contributing thoughtful information about the efficacy and safety of the new booster.<br><br>Your task is to talk with ChatGPT to gather information about the new COVID-19 booster. Your goal is to participate in your neighbors' conversation, providing your insights into this issue and addressing any concerns they might have. |
| | Science | **[Climate change]**<br>July 2024 has officially been recorded as the hottest ever, and your neighbors are concerned about the rising temperatures and future impacts. Some are skeptical about the urgency of climate change, while others want solutions but are unsure about the facts. You want to join the conversation by contributing thoughtful information about climate change, its causes, and possible solutions.<br><br>Your task is to talk with ChatGPT to gather information about the current climate situation, impacts, and solutions, including the reasons behind July 2024 being the hottest month on record. Your goal is to participate in your neighbors' conversation, providing your insights into this issue. |
| | Policy | **[Immigration]**<br>As the 2024 presidential election approaches, your neighbors are discussing immigration policy. Some think immigration benefits economic growth, while others are concerned about its impact on national security and public resources. You want to join the conversation by providing clear information about the pros and cons of immigration policies.<br><br>Your task is to talk with ChatGPT to gather information about the impacts of immigration policies on the economy, national security, etc. Your goal is to participate in the conversation, offering helpful insights into the benefits and challenges of immigration policies. |
| Non-controversial | Health | **[Artificial sweeteners]**<br>Your neighbors have been discussing the potential health effects of artificial sweeteners. Some are concerned about the safety of consuming them regularly, while others believe they are a good alternative to sugar. You want to join the conversation by providing thoughtful information about the health impacts of artificial sweeteners. |
| | | Your task is to talk with ChatGPT to gather information about the health impacts of artificial sweeteners. Your goal is to participate in the conversation, helping your neighbors understand the safety and potential health risks associated with artificial sweeteners. |
| | Science | **[Microplastics]**<br>Microplastics have become a growing environmental concern, and your neighbors are anxious about their effects on the environment. They aren't sure what they can do with microplastics. You want to join the conversation by sharing information about the environmental impact of microplastics and what can be done to reduce their spread.<br><br>Your task is to talk with ChatGPT to gather information about the environmental impact of microplastics, including how they affect ecosystems and what actions can help reduce them. Your goal is to participate in the conversation, helping your neighbors understand the significance of microplastic pollution and how it can be addressed. |
| | Policy | **[Highway infrastructure]**<br>Your neighbors are discussing interstate highway infrastructure. Some are worried about aging roads and frequent repairs, while others are concerned with the environmental impacts. You want to join the conversation by sharing information about the pros and cons of interstate highway reconstruction and possible solutions.<br><br>Your task is to talk with ChatGPT to gather information about the impacts of infrastructure improvements on safety, traffic congestion, and the economy. Your goal is to participate in the conversation, helping your neighbors better understand this issue. |

**Table S2.** Summary statistics of demographics.

| Demographic Characteristics | | N = 937 | |
|---|---|---|---|
| | | n | % |
| Gender | Female | 473 | 50.8% |
| | Male | 439 | 47.1% |
| | Non-binary | 20 | 2.1% |
| Age, mean (SD) | | 45.1 (15.6) | - |
| Race | Asian | 79 | 8.4% |
| | Black | 149 | 15.9% |
| | Hispanic | 96 | 10.2% |
| | Native Hawaiian or Other Pacific Islander | 9 | 1.0% |
| | White | 656 | 70.0% |
| Education | Some high school or less | 9 | 1.0% |
| | High School/GED | 103 | 11.0% |
| | Some college no degree | 218 | 23.3% |
| | Associates/technical degree | 112 | 12.0% |
| | Bachelor’s degree | 315 | 33.7% |
| | Graduate/professional degree | 178 | 19.0% |
| Income | <$25,000 | 123 | 13.1% |
| | $25,000–$49,999 | 181 | 19.3% |
| | $50,000–$74,999 | 204 | 21.8% |
| | $75,000–$99,999 | 143 | 15.3% |
| | $100,000–$149,999 | 163 | 17.4% |
| | ≥$150,000 | 102 | 10.9% |
| Political affiliation | A strong Democrat | 257 | 27.5% |
| | A not very strong Democrat | 115 | 12.3% |
| | Independent, lean toward Democrat | 149 | 15.9% |
| | Independent (close to neither party) | 137 | 14.6% |
| | Independent, lean toward Republican | 91 | 9.7% |
| | A not very strong Republican | 82 | 8.8% |
| | A strong Republican | 105 | 11.2% |

**Table S3.** ChatGPT instruction for annotating prompting strategies.

You are an AI agent tasked with classifying the prompting strategies that users employ when engaging in a hypothetical neighborhood discussion scenario. Users seek information on a single assigned topic by submitting multiple prompts to a conversational AI. Your task is to determine which prompting strategy (if any) is used in each prompt. Output your classification result (e.g., 0, 1, 2) without explanation or additional text.

Dataset Structure:

- ResponseId: Unique identifier for each user.
- Condition: The assigned context for which the user is seeking information. Possible contexts include: climate change, microplastics, COVID-19 booster, artificial sweeteners, immigration, or highway reconstruction.
- Message: The user's prompt.

For each prompt in the dataset, determine which strategy (if any) is used and assign the appropriate code:

| | |
|---|---|
| 0. No specific strategies used | The prompt is a general question or statement without clear strategic structuring. |
| 1. Request for references, relevant studies, statistics, links, or resources | The user explicitly asks for sources, citations, or additional resources.<br>Example: “What about some resources for people if they want to help?” |
| 2. Request to adopt a persona | The user asks the AI to take on a specific role or speak from a particular perspective.<br>Example: “I want you to act like an economics expert and tell me the pros and cons of the current US immigration policy.” |
| 3. Use of delimiters to separate distinct parts of the input | The user clearly marks different sections of their prompt using symbols such as quotes, brackets, or other delimiters.<br>Example: “what are some specific solutions that involve policy change rather than cosmetic """"""""reusable straws""""""""""""ideas” |
| 4. Specification of steps required to complete a task | The user explicitly asks for step-by-step reasoning or instructions.<br>Example: “Can you break this down into steps?” |
| 5. Provision of examples | The user includes examples to guide the AI’s response.<br>Example:<br>“Here’s an example of what I mean: [example]. Can you provide something similar?” |
| 6. Specify the desired length of the output | The user sets a constraint on response length.<br>Example: “Summarize in 100 words or fewer how to convince someone that the new Covid-19 booster is worth getting.” |
| 7. Specify the desired communication styles, formats, or tones of the output | The user requests a specific format, structure, style, or tone for the response. This includes using bullet points, lists, simplified language, conversational tone, or any other specific presentation style.<br>Example: “What are microplastics? use bullet points and be brief” |
| 8. Provide contextual information | The user provides contextual information where and how the information will be used, especially when it comes to hypothetical neighborhood discussion scenarios.<br>Example: “How can I persuade my neighbors to get the boosters” |

**Table S4.** Measures of issue-specific pre-existing attitude and knowledge.

| **Items** | **Response** |
|---|---|
| *Issue-specific pre-existing attitude*<br>1. How effective do you think the COVID-19 vaccines are?<br>2. How severe do you think the environmental impacts of climate change are?<br>3. How severe do you think the environmental impacts of microplastics are?<br>4. How beneficial do you believe immigration is for American society?<br>5. How beneficial do you think highway reconstruction is for American society?<br>6. How safe do you believe artificial sweeteners are for human health? | <br>1. 1 = not effective at all, 4 = very effective<br>2. 1 = not severe at all, 4 = very severe<br>3. 1 = not severe at all, 4 = very severe<br>4. 1 = very harmful, 4 = very beneficial<br>5. 1 = very harmful, 4 = very beneficial<br>6. 1 = not safe at all, 4 = very safe |
| Issue-specific pre-existing knowledge<br>1. How would you rate your level of knowledge about the following topics? | <br>1. 1 = very little or no knowledge, 4 = extensive knowledge |

**Table S5.** Measures of AI familiarity.

| Question | Responses |
|---|---|
| In your opinion, which of the following technologies, if any, uses AI? | 1. Virtual assistants (e.g., Siri, Google Assistant, Amazon Alexa)<br>2. Smart speakers (e.g., Amazon Echo, Google Home, Apple Homepod)<br>3. Smart health/medical sensors (e.g., smart weight scale, smart blood pressure monitors, smart blood sugar/glucose meter)<br>4. Chatbots/robots that interact with humans (e.g., social robots, custom service bots, Replika)<br>5. Facebook photo tagging<br>6. Google Search<br>7. Content recommendations (e.g., recommendations for Netflix movies, Amazon products/ebooks, social media feeds)<br>8. Google Translate<br>9. Self-driving cars and trucks / auto-driving vehicles (e.g., Tesla)<br>10. Industrial robots used in manufacturing<br>11. Drones that do not require a human controller<br>12. Staffless stores (e.g., Amazon Go)<br>13. None of the above |

**Table S6.** Measures of issue-specific attitude (post-experiment). (R) means the values for that item is reversely coded.

| **Items** | **Response** |
|---|---|
| *COVID-19 vaccine safety*<br>1. COVID-19 vaccines are safe.<br>2. COVID-19 vaccines contain dangerous ingredients. (R)<br>3. COVID-19 vaccines are effective at preventing COVID-19 infection.<br>4. COVID-19 vaccines are unnecessary since COVID-19 is harmless. (R) | 1 = strongly disagree, 5 = strongly agree |
| *Climate change severity*<br>1. How concerned are you about the impacts of climate change?<br>2. In your judgment, how likely are you to experience serious threats to your health as a result of climate change?<br>3. In your judgment, how likely do you think it is that climate change will have very harmful, long-term impacts on our society?<br>4. How serious of a threat do you think climate change is to the natural environment? | 1 = not concerned at all, 5 = very concerned<br>1 = very unlikely, 5 = very likely<br>1 = not serious at all, 5 = very serious |
| *Immigration benefits*<br>1. In general, immigrants who come to the United States today help the country and make it a better place to live.<br>2. Immigrants take jobs away from American workers. (R)<br>3. Immigration increases the rates of crime. (R)<br>4. Immigration contributes to economic growth in the United States. | 1 = strongly disagree, 5 = strongly agree |
| *Artificial sweetener safety*<br>1. Artificial sweeteners are harmful to human health. (R)<br>2. It does not bother me if my foods contain artificial sweeteners.<br>3. I am worried about what effects artificial sweeteners could have on my body. (R)<br>4. Artificial sweeteners are completely safe. | 1 = strongly disagree, 5 = strongly agree |
| *Microplastics severity*<br>1. How concerned are you about the health impacts of microplastics?<br>2. How concerned are you about the impacts of microplastics on the environment?<br>3. How concerned are you about the impacts of microplastics on animals and plant growth?<br>4. How concerned are you about microplastics accumulating in the human body? | 1 = not concerned at all, 5 = very concerned |
| *Highway reconstruction benefits*<br>1. Highway reconstruction does not help reduce traffic congestion in the long term. (R)<br>2. Highway reconstruction stimulates long-term economic growth.<br>3. Highway reconstruction significantly increases pollution levels. (R)<br>4. Highway reconstruction improves road safety. | 1 = strongly disagree, 5 = strongly agree |

**Table S7.** Measures of perceived AI response quality and interaction quality.

| Items | Response |
|---|---|
| *AI response perception*<br>What do you think about ChatGPT's responses using the scale below? | 1. Generic - In-depth<br>2. Rich information - Lack of information (R)<br>3. Accurate - Inaccurate (R)<br>4. Strong reasoning - Poor reasoning (R)<br>5. Easy to understand - Hard to understand (R)<br>6. Too detailed - Too short (R)<br>7. Irrelevant - Relevant<br>8. Useful - Useless (R)<br>9. Complete - Incomplete (R)<br>10. Biased - Unbiased<br>11. Clear - Ambiguous (R)<br>12. Outdated - Up-to-date |
| *AI interaction perception*<br>To what extent do you agree or disagree with the following statements?<br>(1 = strongly disagree, 7 = strongly agree) | 1. In this conversation, ChatGPT's responses did not follow my instructions. (R)<br>2. In this conversation, ChatGPT made up information that couldn't be verified from reliable sources. (R)<br>3. In this conversation, ChatGPT agreed with my opinion too much. (R)<br>4. In this conversation, ChatGPT's responses included unlawful, unethical, harmful, or biased content. (R)<br>5. In this conversation, ChatGPT was able to understand my questions and instructions clearly.<br>6. Overall, I am satisfied with the quality of this conversation with ChatGPT.<br>7. I would like to use ChatGPT again for similar tasks. |

**Table S8.** Measures of AI perceptions in terms of perceived likability, trustworthiness, and intelligence.

| **Dimensions** | **Items** |
|---|---|
| Perceived likability | 1. Not likable - Likable<br>2. Unpleasant - Pleasant<br>3. Unappealing - Appealing<br>4. Irritating - Not irritating |
| Perceived trustworthiness | 5. Not credible - Credible<br>6. Untrustworthy - Trustworthy<br>7. Dishonest - Honest<br>8. Unreliable - Reliable<br>9. Biased - Unbiased<br>10. Insincere - Sincere |
| Perceived intelligence | 11. Unintelligent - Intelligent<br>12. Dumb - Smart<br>13. Incapable - Capable |

**Table S9.** Summary statistics of key dependent variables across issue topics.

| **Controversy** | | Controversial | | | Non-controversial | | |
|---|---|---|---|---|---|---|---|
| **Issue** | | *Health* | *Science* | *Policy* | *Health* | *Science* | *Policy* |
| **Topic** | | COVID-19 booster | Climate change | Immigration | Artificial sweeteners | Microplastics | Highway infrastructure |
| Pre-knowledge (1-4) | M | 3.29 | 3.17 | 3.18 | 2.73 | 2.54 | 2.21 |
| | SD | 0.70 | 0.84 | 0.96 | 0.96 | 0.94 | 0.99 |
| Pre-attitude (1-4) | M | 2.98 | 3.36 | 2.78 | 2.24 | 3.35 | 3.44 |
| | SD | 1.10 | 0.76 | 0.93 | 0.74 | 0.68 | 0.65 |
| Post-attitude (1-5) | M | 3.91 | 4.07 | 3.59 | 2.70 | 4.04 | 3.89 |
| | SD | 1.15 | 0.96 | 1.08 | 1.03 | 0.84 | 0.64 |
| Perceived response quality (1-7) | M | 5.60 | 5.87 | 5.55 | 5.95 | 5.86 | 5.87 |
| | SD | 0.92 | 0.71 | 0.89 | 0.71 | 0.78 | 0.83 |
| Perceived interaction quality (1-7) | M | 4.29 | 4.44 | 4.17 | 4.35 | 4.41 | 4.35 |
| | SD | 0.63 | 0.42 | 0.58 | 0.45 | 0.47 | 0.44 |
| Perceived AI likability (1-7) | M | 5.54 | 5.82 | 5.59 | 5.73 | 5.77 | 5.67 |
| | SD | 1.17 | 1.09 | 1.01 | 1.15 | 1.15 | 1.22 |
| Perceived AI credibility (1-7) | M | 5.41 | 5.72 | 5.28 | 5.66 | 5.72 | 5.49 |
| | SD | 1.10 | 0.95 | 1.20 | 0.99 | 1.04 | 1.02 |
| Perceived AI intelligence (1-7) | M | 5.81 | 6.26 | 5.78 | 6.07 | 6.17 | 6.11 |
| | SD | 1.14 | 0.85 | 1.07 | 1.00 | 0.97 | 0.96 |

**Table S10.** Regression coefficients and standard errors predicting the count of (A) prompting strategy and three prompting strategy categories, respectively: (B) user-supplied information, (C) style-related requests, and (D) content-related requests.

| | (A) prompting strategy | (B) user-supplied information | (C) style-related requests | (D) content-related requests |
|---|---|---|---|---|
| Gender: Female | -0.29 (0.15) | 0.04 (0.24) | -0.35 (0.25) | -0.59 (0.27)* |
| Age | -0.01 (0.01) | -0.02 (0.01)* | 0.00 (0.01) | 0.00 (0.01) |
| Race: White | 0.29 (0.18) | 0.21 (0.27) | 0.14 (0.30) | 0.53 (0.33) |
| Education | 0.16 (0.07)* | 0.15 (0.10) | 0.14 (0.11) | 0.19 (0.12) |
| Income | 0.07 (0.05) | 0.00 (0.08) | 0.03 (0.08) | 0.20 (0.09)* |
| Political affiliation | -0.12 (0.04)** | -0.12 (0.07) | -0.05 (0.07) | -0.19 (0.08)* |
| Issue: controversial | 0.16 (0.17) | -0.04 (0.26) | 0.03 (0.28) | 0.47 (0.29) |
| Issue: policy | -0.16 (0.19) | -0.12 (0.30) | 0.08 (0.31) | -0.39 (0.34) |
| Issue: health | 0.19 (0.19) | 0.26 (0.29) | 0.36 (0.32) | 0.05 (0.32) |
| Pre-knowledge | -0.26 (0.09)** | -0.15 (0.14) | -0.37 (0.15)* | -0.33 (0.16)* |
| Pre-attitude | 0.11 (0.10) | 0.21 (0.15) | 0.31 (0.17) | -0.17 (0.17) |
| AI familiarity | 0.04 (0.03) | 0.03 (0.04) | 0.10 (0.05)* | 0.00 (0.04) |
| Constant | -1.41 (0.57)* | -2.32 (0.88)** | -3.64 (0.95)*** | -2.06 (0.98)* |
| R2 | 0.095 | 0.058 | 0.081 | 0.123 |

∗∗∗ $p < 0.001$, ∗∗ $p < 0.01$, ∗ $p < 0.05$

**Table S11.** Regression coefficients and standard errors of issue controversy, prompting strategy count, and other factors predicting ChatGPT's citation count, structure count, and 5 communication styles.

| | **Citation Count** | **Structure Count** | **Cognitive complexity** | **Self-revealing** | **Information-seeking** | **Fact-oriented** | **Action-seeking** |
|---|---|---|---|---|---|---|---|
| Issue: controversial | 1.05 (0.11)*** | -5.84 (0.70)*** | 0.06 (0.01)*** | 0.03 (0.02) | 0.01 (0.02) | 0.00 (0.00) | 0.08 (0.02)*** |
| Prompting strategy | -0.04 (0.09) | 0.54 (0.55) | 0.00 (0.01) | 0.00 (0.01) | -0.03 (0.01) | 0.00 (0.00) | 0.01 (0.02) |
| Issue: policy | -0.57 (0.15)*** | -0.79 (0.91) | -0.02 (0.01) | 0.04 (0.02)* | -0.01 (0.02) | -0.01 (0.00) | 0.01 (0.03) |
| Issue: health | -0.03 (0.15) | -3.47 (0.92)*** | -0.05 (0.01)*** | 0.01 (0.02) | -0.04 (0.02) | 0.00 (0.00) | 0.21 (0.03)*** |
| Pre-attitude | 0.10 (0.06) | -0.26 (0.40) | 0.00 (0.00) | 0.00 (0.01) | -0.01 (0.01) | 0.00 (0.00) | 0.05 (0.01)*** |
| User: cognitive complexity | 0.40 (0.40) | 4.37 (2.49) | 0.08 (0.02)*** | -0.06 (0.06) | 0.05 (0.06) | 0.02 (0.01) | -0.05 (0.08) |
| User: self-revealing | -0.21 (0.13) | -3.11 (0.82)*** | -0.04 (0.01)*** | 0.07 (0.02)*** | 0.11 (0.02)*** | 0.00 (0.00) | 0.07 (0.03)** |
| User: information-seeking | -0.24 (0.35) | 0.58 (2.17) | 0.08 (0.02)*** | -0.20 (0.05)*** | -0.19 (0.05)*** | 0.05 (0.01)*** | -0.02 (0.07) |
| User: fact-oriented | -0.54 (0.21)** | 0.16 (1.29) | -0.01 (0.01) | -0.02 (0.03) | -0.02 (0.03) | 0.01 (0.00) | -0.04 (0.04) |
| User: action-seeking | 0.20 (0.22) | -1.18 (1.38) | -0.02 (0.01) | 0.04 (0.03) | 0.02 (0.03) | 0.01 (0.01) | 0.08 (0.05) |
| User: word count | 0.00 (0.00) | -0.02 (0.00)*** | 0.00 (0.00) | 0.00 (0.00) | 0.00 (0.00) | 0.00 (0.00) | 0.00 (0.00) |
| AI: word count | 0.00 (0.00)** | 0.01 (0.00)*** | 0.00 (0.00)* | 0.00 (0.00) | 0.00 (0.00)*** | 0.00 (0.00) | 0.00 (0.00) |
| Constant | 0.46 (0.50) | 8.49 (3.07)** | 0.45 (0.03)*** | 0.24 (0.07)*** | 0.31 (0.08)*** | 0.94 (0.01)*** | -0.07 (0.10) |
| R2 | 0.152 | 0.559 | 0.236 | 0.058 | 0.087 | 0.049 | 0.120 |

∗∗∗ $p < 0.001$, ∗∗ $p < 0.01$, ∗ $p < 0.05$

**Table S12.** Regression coefficients and standard errors of issue controversy, three categories of prompting strategies, and other factors predicting ChatGPT's citation count, structure count, and 5 communication styles.

| | **Citation Count** | **Structure Count** | **Cognitive complexity** | **Self-revealing** | **Information-seeking** | **Fact-oriented** | **Action-seeking** |
|---|---|---|---|---|---|---|---|
| Issue: controversial | 1.04 (0.11)*** | -5.77 (0.70)*** | 0.06 (0.01)*** | 0.02 (0.02) | 0.01 (0.02) | 0.00 (0.00) | 0.08 (0.02)*** |
| Prompting strategy: user-supplied information | 0.12 (0.15) | 1.77 (0.94) | -0.01 (0.01) | 0.02 (0.02) | -0.02 (0.02) | 0.00 (0.00) | 0.08 (0.03)** |
| Prompting strategy: style-related requests | -0.25 (0.17) | 0.93 (1.06) | -0.03 (0.01)** | -0.04 (0.02) | -0.05 (0.03) | 0.00 (0.00) | -0.02 (0.04) |
| Prompting strategy: content-related requests | -0.02 (0.14) | -0.67 (0.86) | 0.03 (0.01)** | 0.03 (0.02) | -0.02 (0.02) | 0.00 (0.00) | -0.03 (0.03) |
| Issue: policy | -0.57 (0.15)*** | -0.82 (0.91) | -0.01 (0.01) | 0.05 (0.02)* | -0.01 (0.02) | -0.01 (0.00) | 0.01 (0.03) |
| Issue: health | -0.03 (0.15) | -3.50 (0.92)*** | -0.05 (0.01)*** | 0.01 (0.02) | -0.04 (0.02) | 0.00 (0.00) | 0.20 (0.03)*** |
| Pre-attitude | 0.10 (0.06) | -0.31 (0.40) | 0.00 (0.00) | 0.01 (0.01) | 0.00 (0.01) | 0.00 (0.00) | 0.05 (0.01)*** |
| User: cognitive complexity | 0.43 (0.40) | 4.53 (2.49) | 0.08 (0.02)*** | -0.06 (0.06) | 0.05 (0.06) | 0.02 (0.01) | -0.04 (0.08) |
| User: self-revealing | -0.23 (0.13) | -3.32 (0.82)*** | -0.04 (0.01)*** | 0.07 (0.02)*** | 0.11 (0.02)*** | 0.00 (0.00) | 0.06 (0.03)* |
| User: information-seeking | -0.23 (0.35) | 0.47 (2.16) | 0.08 (0.02)*** | -0.20 (0.05)*** | -0.19 (0.05)*** | 0.05 (0.01)*** | -0.02 (0.07) |
| User: fact-oriented | -0.56 (0.21)** | -0.02 (1.30) | -0.01 (0.01) | -0.02 (0.03) | -0.02 (0.03) | 0.01 (0.00) | -0.05 (0.04) |
| User: action-seeking | 0.21 (0.22) | -1.34 (1.38) | -0.01 (0.01) | 0.04 (0.03) | 0.02 (0.03) | 0.01 (0.01) | 0.08 (0.05) |
| User: word count | 0.00 (0.00) | -0.02 (0.00)*** | 0.00 (0.00) | 0.00 (0.00) | 0.00 (0.00) | 0.00 (0.00) | 0.00 (0.00) |
| AI: word count | 0.00 (0.00)** | 0.01 (0.00)*** | 0.00 (0.00) | 0.00 (0.00) | 0.00 (0.00)*** | 0.00 (0.00) | 0.00 (0.00) |
| Constant | 0.47 (0.50) | 8.64 (3.07) | 0.45 (0.03)*** | 0.24 (0.07)** | 0.31 (0.08)*** | 0.94 (0.01)*** | -0.06 (0.10) |
| R2 | 0.153 | 0.560 | 0.256 | 0.062 | 0.086 | 0.047 | 0.128 |

∗∗∗ $p < 0.001$, ∗∗ $p < 0.01$, ∗ $p < 0.05$

**Table S13.** Regression coefficients and standard errors of ChatGPT responses and other factors predicting users' perceptions of ChatGPT and its responses, and issue-relevant post attitudes.

| | **Perceived response quality** | **Perceived interaction quality** | **Perceived AI likability** | **Perceived AI credibility** | **Perceived AI intelligence** | **Issue-relevant post attitude** |
|---|---|---|---|---|---|---|
| Citation Count | 0.02 (0.02) | 0.01 (0.01) | -0.03 (0.03) | 0.00 (0.03) | 0.00 (0.03) | 0.00 (0.02) |
| Structure Count | 0.00 (0.00) | 0.00 (0.00) | 0.00 (0.00) | 0.00 (0.00) | 0.00 (0.00) | 0.00 (0.00) |
| Cognitive complexity | -1.19 (0.35)*** | -0.20 (0.22) | -1.07 (0.50)* | -0.75 (0.45) | -0.95 (0.44)* | 0.70 (0.32)* |
| Self-revealing | 0.05 (0.15) | 0.15 (0.09) | 0.13 (0.21) | 0.22 (0.19) | -0.05 (0.19) | 0.01 (0.13) |
| Information-seeking | -0.05 (0.13) | -0.05 (0.08) | 0.03 (0.19) | 0.03 (0.17) | -0.07 (0.17) | 0.06 (0.12) |
| Fact-oriented | 0.08 (0.80) | 0.27 (0.50) | -0.28 (1.16) | -0.62 (1.05) | -0.40 (1.03) | -0.53 (0.74) |
| Action-seeking | -0.03 (0.10) | 0.00 (0.06) | 0.02 (0.14) | -0.14 (0.12) | 0.04 (0.12) | 0.12 (0.09) |
| Issue: controversial | -0.16 (0.06)* | -0.08 (0.04)* | -0.04 (0.09) | -0.15 (0.08) | -0.11 (0.08) | 0.22 (0.06)*** |
| Issue: policy | -0.11 (0.07) | -0.14 (0.05)** | -0.17 (0.10) | -0.29 (0.09)** | -0.25 (0.09)** | -0.10 (0.07) |
| Issue: health | 0.01 (0.08) | -0.04 (0.05) | -0.18 (0.11) | -0.05 (0.10) | -0.22 (0.10)* | -0.18 (0.07)* |
| Pre-attitude | 0.24 (0.04)*** | 0.12 (0.02)*** | 0.14 (0.05)** | 0.25 (0.05)*** | 0.17 (0.05)*** | 0.76 (0.03)*** |
| AI familiarity | 0.02 (0.01) | 0.01 (0.01)* | 0.03 (0.01)* | 0.01 (0.01) | 0.01 (0.01) | -0.01 (0.01) |
| Gender: Female | 0.09 (0.06) | 0.04 (0.04) | 0.03 (0.08) | 0.01 (0.08) | 0.05 (0.07) | 0.10 (0.05) |
| Age | 0.01 (0.00)** | 0.01 (0.00)*** | 0.01 (0.00)*** | 0.01 (0.00)*** | 0.01 (0.00)** | 0.00 (0.00) |
| Race: White | -0.08 (0.07) | -0.05 (0.04) | -0.19 (0.10)* | -0.21 (0.09)* | -0.13 (0.09) | 0.09 (0.06) |
| Education | -0.05 (0.02)* | -0.04 (0.01)** | -0.05 (0.03) | -0.05 (0.03) | -0.03 (0.03) | 0.01 (0.02) |
| Political affiliation | -0.02 (0.02) | -0.02 (0.01) | 0.00 (0.02) | -0.01 (0.02) | 0.03 (0.02) | -0.07 (0.01)*** |
| Constant | 5.61 (0.83)*** | 3.77 (0.52)*** | 6.04 (1.20)*** | 5.84 (1.08)*** | 6.34 (1.07)*** | 1.85 (0.76)* |
| R2 | 0.110 | 0.097 | 0.037 | 0.082 | 0.037 | 0.561 |

*** $p < 0.001$, ** $p < 0.01$, * $p < 0.05$